\documentclass[a4paper,12pt]{article}

\usepackage{fabdoc,fabmacromath}

\newtheorem{lemma}{Lemma}[section]
\newtheorem{theorem}{Theorem}[section]
\newtheorem{corollary}{Corollary}[section]

\newcommand{\be}{\begin{equation}}
  \newcommand{\ee}{\end{equation}}
\newcommand{\bea}{\begin{eqnarray}}
  \newcommand{\eea}{\end{eqnarray}}
\newcommand{\sh}{\sinh}
\newcommand{\ch}{\cosh}
\newcommand{\prf}{\noindent{\bf Proof}\ }
\newcommand{\qed}{{\hfill $\Box$}}

\newcommand{\proof}{{\bfseries Proof.}}

\begin{document}
\begin{center}

  {\Large\bfseries Propagators for Noncommutative Field Theories}

  \vskip 4ex

  Razvan \textsc{Gurau}, 
  Vincent \textsc{Rivasseau} and
  Fabien \textsc{Vignes-Tourneret}

  \vskip 3ex  

  \textit{Laboratore de Physique Th\'eorique, B\^at.\ 210, 
    Universit\'e Paris XI \\ F-91405 Orsay Cedex, France}
  \\
  e-mail: \texttt{razvan.gurau@th.u-psud.fr}, 
  \texttt{vincent.rivasseau@th.u-psud.fr}, 
  \texttt{fabien.vignes@th.u-psud.fr}
\end{center}

\vskip 5ex

\begin{abstract}
  In this paper we provide exact expressions for propagators of noncommutative
  Bosonic or Fermionic field theories after adding terms of the Grosse-Wulkenhaar
  type in order to ensure Langmann-Szabo covariance. We emphasize 
  the new Fermionic case and we give in particular all necessary bounds for the multiscale
  analysis and renormalization of the noncommutative Gross-Neveu model. 
\end{abstract}

%\tableofcontents

\section{Introduction}

This paper is the first of a series in which we plan to extend the proof of perturbative 
renormalizability of noncommutative $\phi^4_4$ field theory
\cite{GrWu03-1,GrWu04-3,Rivasseau2005bh} to other noncommutative models 
(see \cite{DouNe} for a general review on noncommutative field theories).

We have in mind in particular Fermionic field theories either of the relativistic type,
such as the Gross-Neveu model in Euclidean two dimensional space \cite{Mitter:1974cy,Gross:1974jv}, 
or of the type used in condensed matter for many body theory, in
which there is no symmetry between time and space. In the commutative case, these non-relativistic theories 
are just renormalizable in any dimension \cite{FT,Benfatto:1996ngIN,Salmhofer}. 
Their noncommutative version should be relevant 
for the study of Fermions in 2 dimensions in magnetic fields, hence for the quantum Hall effect.
Of course a future goal is also to find the right extension of the Grosse-Wulkenhaar method to
gauge theories. 

In this paper we generalize the computation of the Bosonic $\phi^4_4$ propagator
of \cite{GrWu04-3} and provide the exact expression of the propagators of Fermionic 
noncommutative field theories on the Moyal plane. We will restrict our analysis to one pair 
of noncommuting coordinates, as the generalization to several pairs are 
trivial. These propagators are not the ordinary commutative propagators: 
they have to be modified to obey Langmann-Szabo duality, according to the 
pioneering papers \cite{LaSz,GrWu04-3}. We propose to call {\it vulcanization}
this modification of the theory, and to call {\it vulcanized} the resulting theory and propagators.

Like in \cite{Rivasseau2005bh} we can slice the corresponding noncommutative heat kernels 
according to the Schwinger parameters in order to derive a multiscale analysis. In this framework
the  theory with a finite number of slices has a cutoff, and the removal of the cutoff 
(ultraviolet limit) corresponds to summing over infinitely many slices. 
However for the Fermionic propagators treated in this paper 
this multiscale analysis is harder than in the Bosonic case.
In $x$-space the propagator end-terms oscillate rather than decay as in 
the  Bosonic $\phi^4_4$ case. In matrix basis the behavior of the propagator
is governed by a non-trivial critical point in parameter space and in contrast
with the Bosonic case there is no general scaled decay for all indices.
The main result of this paper is the detailed analysis of this critical point,
leading to Theorem \ref{maintheorem}, namely to the bounds required for the multiscale analysis of the 
2 dimensional Euclidean Non-Commutative Gross-Neveu model. 
These bounds are slightly worse than in the $\phi^4_4$ case. 
However they should suffice for a complete proof of renormalizability of the model 
to all orders (see the discussion after Theorem \ref{maintheorem}). This complete proof
is postponed to a future paper \cite{RenNCGN05}.

\section{Conventions}
\setcounter{equation}{0}

The two dimensional Moyal space ${\mathbb R}^2_\theta$
is defined by the following associative non-commutative star product
\begin{equation}
  (a\star b)(x) = \int \frac{d^2k}{(2\pi)^2} \int d^2 y \; a(x{+}\tfrac{1}{2}
  \theta {\cdot} k)\, b(x{+}y)\, \mathrm{e}^{\mathrm{i} k \cdot y}\;.
\label{starprod}
\end{equation}
which corresponds to a constant commutator:
\begin{equation}
  [ x^i , x^j ]  = i  \Theta ^{ij} \ \ ,  {\rm where } 
 \ \Theta=\begin{pmatrix}\phantom{-}0&\theta\\-\theta&0\end{pmatrix}. 
\end{equation}

In order to perform the second quantization one must first identify 
the Hilbert space of states of the first quantization. If we deal with a 
real field theory (for instance the $\phi^4$ theory which is treated in 
detail in \cite{GrWu03-2}) one considers a real Hilbert space, 
whereas for a complex field theory (e.g. any Fermionic field theory) one has
to use a complex Hilbert space.

A basis in this Hilbert space is given by the functions 
$f_{mn}$ defined in \cite{Gracia-Bondia1987kw,GrWu03-2}.  

Any complex-valued function defined on the plane can be decomposed in 
this basis as:
\begin{equation}
  \chi(x)=\sum_{m,n} \chi_{mn}f_{mn}(x).
\end{equation}

A crucial observation is that $\bar{f}_{mn}(x)=f_{nm}(x)$ (which can 
be verified on the explicit expression for $f_{mn}$). A real function 
in this basis obeys:
\begin{equation}
  \bar{\chi}(x)=\chi(x)\Rightarrow\overline{\sum\chi_{mn}f_{mn}(x)}=
  \sum\chi_{pq}f_{pq(x)}\Rightarrow\bar{\chi}_{mn}=\chi_{nm} .
\end{equation}

The scalar product (which must be sesquilinear for a 
complex Hilbert space) is then defined as:
\begin{eqnarray}
  \langle\phi,\chi\rangle&=&\int\frac{d^2x}{2\pi\theta}
  ~\bar{\phi}(x)\chi(x)=\sum\bar{\phi}_{pq}
  \chi_{rs}\int\frac{d^2x}{2\pi\theta}\bar{f}_{pq}f_{rs}\nonumber\\
  &=&
  \sum\bar{\phi}_{pq}\chi_{rs}\delta_{pr}\delta_{qs}=\sum\bar{\phi}_{pq}\chi_{pq} .
\end{eqnarray}

Notice that if $\phi$ is a real field we have:
\begin{equation}
  \langle\phi,\chi\rangle=\sum\phi_{qp}\chi_{pq},
\end{equation}
so that our conventions restrict to those in \cite{GrWu03-2} for 
the real $\phi^4$ theory.
With this convention $\langle f_{kl},\chi\rangle=\chi_{mn}\int \bar{f}_{kl}f_{mn}=\chi_{kl}$.
A linear operator on this space acts like:
\begin{equation}
  [A\phi]_{kl}=\langle f_{kl},A\phi\rangle=\sum_{m,n}\langle f_{kl},Af_{mn}\rangle\phi_{mn}.
\end{equation}

At this point the convention consistent with that for the real 
$\phi^4$ theory found in \cite{GrWu03-2} is to note:
\begin{equation}
  \langle f_{kl},Af_{mn}\rangle=\int\frac{d^2x}{2\pi\theta}\bar{f}_{kl}(x)
  \int d^2y A(x,y)f_{mn}(y):=A_{l,k;m,n} .
\end{equation}
With this convention the product of operators is
\begin{equation}
  \lsb AB\rsb_{p,q;r,s}=\langle f_{qp},ABf_{rs}\rangle=\sum_{t,u}\langle f_{qp},Af_{tu}\rangle
  \langle f_{tu},Bf_{rs}\rangle=\sum_{t,u}A_{p,q;t,u}B_{u,t;r,s}
\end{equation}
and the identity operator $I$ has the matrix elements:
\begin{equation}
  \phi_{mn}=\sum_{p,q}\langle f_{mn},If_{pq}\rangle\phi_{pq}
  \Rightarrow I_{n,m;p,q}=\delta_{pm}\delta_{nq}.
\end{equation}

We pass now to the second quantization.
The quadratic part of the action is generically (in $x$ space):
\begin{equation}
  S=\int d^2x~\bar{\chi}(x)~H(x,y)~\chi(y) .
\end{equation}

In the matrix basis one has the action:
\begin{equation}
  S=2\pi\theta\sum_{m,n,k,l}\frac{1}{2} \bar{\chi}_{pq}\Big{(}2\int\frac{d^2x}{2\pi\theta}\bar{f}_{pq}(x)
  H(x,y)f_{kl}(y)\Big{)} \chi_{kl}.
\end{equation}
We define the Hamiltonian in the matrix basis as:
\begin{equation}
  H_{q,p;k,l}=2\int\frac{d^2x}{2\pi\theta}\bar{f}_{pq}(x)
  H(x,y)f_{kl}(y)
\end{equation}
where the $2$ has been included in order to maintain the conventions 
in \cite{GrWu03-2} for the case of a real field.

\section{Non-Commutative Schwinger Kernels}
\setcounter{equation}{0}

In this section we provide the explicit formulas for the Schwinger representation
of the non-commutative kernels or propagators of free scalar Bosonic and 
spin 1/2 Fermionic theories on the Moyal plane. These formulas are 
essential for a multiscale analysis based on slicing the Schwinger parameter.

The different propagators of interest are expressed  
via the Schwinger parameter trick as:
\begin{equation}
  H^{-1}=\int_0^{\infty}dt~e^{-tH} .
\end{equation} 

\subsection{Bosonic $\boldsymbol{x}$ Space Kernel}

We define $x\wedge x'=x_0x'_1-x_1x'_0$ and $x\cdot x'=x_0x'_0+x_1x'_1$.
The following lemma generalizes the Mehler kernel \cite{simon79:funct}:
\begin{lemma}
  \label{HinXspace}Let H be:
  \begin{equation}
    H=\frac{1}{2}\big{[}-\partial_0^2-\partial_1^2+
    \Omega^2x^2-2\imath B(x_0\partial_1-x_1\partial_0)\Big{]}.
  \end{equation}     
  The integral kernel of the operator $e^{-tH}$ is:
  \begin{equation}
    e^{-tH}(x,x')=\frac{\Omega}{2\pi\sh\Omega t}e^{-A},
  \end{equation}
  \begin{equation}
    A=\frac{\Omega\ch\Omega t}{2\sh\Omega t}(x^2+x'^2)-
    \frac{\Omega\ch Bt}{\sh\Omega t}x\cdot x'-\imath
    \frac{\Omega\sh Bt}{\sh\Omega t}x\wedge x'.
  \end{equation}
\end{lemma}
\prf
We note that the kernel is correctly normalized: as    
$\Omega =B\rightarrow 0$ we  
have
\begin{equation}
  e^{-tH}(x,x')\rightarrow\frac{1}{2\pi t}e^{-\frac{|x-x'|^2}{2t}},
\end{equation}
which is the normalized heat kernel.\\
We must then check the equation
\begin{equation}\label{diffbasic}
  \frac{d}{dt}e^{-tH}+He^{-tH}=0.
\end{equation}
In fact
\begin{eqnarray}
  \frac{d}{dt}e^{-tH}&=&\frac{\Omega e^{-A}}{2\pi\sh\Omega t}
  \Big{\{}-\Omega\coth\Omega t +
  \frac{\Omega^2}{2\sh^2\Omega t}(x^2+x'^2)
  \nonumber\\
  &+&\Omega\frac{B\sh\Omega t\sh Bt-\Omega\ch\Omega t\ch Bt}
  {\sh^2\Omega t}x\cdot x' 
  \nonumber\\
  &+&\imath\Omega\frac{B\ch Bt\sh\Omega t-\Omega\sh Bt\ch\Omega t}
  {\sh^2\Omega t}x\wedge x'
  \Big{\}}.
\end{eqnarray}
Moreover
\begin{eqnarray}  
  \frac{(-\partial^2_1-\partial^2_2)}{2}e^{-tH}
  &=& \frac{\Omega e^{-A}}{2\pi\sh\Omega t}\Big{\{}
  \Omega\coth\Omega t -\frac{1}{2}
  \Big{[}\frac{\Omega\ch\Omega t}{\sh\Omega t}x-\frac{\Omega\ch 
    Bt}{\sh\Omega t}x'\Big{]}^2
  \nonumber\\
  &+&\frac{\Omega^2\sh^2 Bt}{2\sh^2\Omega t}x'^2
  +\imath\frac{\Omega^2\ch\Omega t\sh Bt}{\sh^2\Omega t}x\wedge x'
  \Big{\}}
\end{eqnarray}
and
\begin{equation}
  \imath B(x_1\partial_2-x_2\partial_1)= \frac{\Omega}{2\pi\sh\Omega 
    t}e^{-A}\Big{\{}
  (-\imath\frac{B\Omega\ch Bt}{\sh\Omega t}x\wedge x'+
  \frac{B\Omega\sh Bt}{\sh\Omega t}x\cdot x')
  \Big{\}} .
\end{equation}
It is now straightforward to verify the 
differential equation (\ref{diffbasic}). \hfill\qed

\begin{corollary}
  Let H be:
  \begin{equation}
    H=\frac{1}{2}\big{[}-\partial_0^2-\partial_1^2+
    \Omega^2x^2-2\imath\Omega(x_0\partial_1-x_1\partial_0)\Big{]}.
  \end{equation}
  The integral kernel of the operator $e^{-tH}$ is:
  \begin{equation}
    e^{-tH}(x,x')=\frac{\Omega}{2\pi\sh\Omega t}e^{-A},
  \end{equation}
  \begin{equation}
    A=\frac{\Omega\ch\Omega t}{2\sh\Omega t}(x^2+x'^2)-
    \frac{\Omega\ch \Omega t}{\sh\Omega t}x\cdot x'-\imath
    \Omega x\wedge x'.
  \end{equation}
\end{corollary}

\subsection{Fermionic $\boldsymbol{x}$ Space Kernel}

The two-dimensional free commutative Fermionic field theory is defined by the 
Lagrangian
\begin{equation}
\label{eq:LGN}
{\cal L}=\psib(x)\lbt\ps+\mu\rbt\psi(x). 
\end{equation}
The propagator of the theory $\lbt\ps + \mu\rbt^{-1}(x,y)$ can be calculated thanks to
the usual heat kernel method as
\begin{align}
  \lbt\ps
  +\mu\rbt^{-1}(x,y)&=\lbt-\ps+\mu\rbt\lbt\lbt\ps+\mu\rbt\lbt-\ps+\mu\rbt\rbt^{-1}(x,y)\nonumber\\
  &=\lbt-\ps+\mu\rbt\lbt p^{2}+\mu^{2}\rbt^{-1}(x,y)\\
  &=\lbt-\ps+\mu\rbt\int_{0}^{\infty}\frac{dt}{4\pi t}\,
  e^{-\frac{(x-y)^{2}}{4t}-\mu^{2}t}\\
  &=\int_{0}^{\infty}\frac{dt}{4\pi
    t}\,\lbt\frac{-\imath}{2t}(\xs-\ys)+\mu\rbt e^{-\frac{(x-y)^{2}}{4t}-\mu^{2}t}.
\end{align}
In the noncommutative case we have to modify the free action, adding a Grosse-Wulkenhaar term to 
implement Langmann-Szabo duality. This shall prevent ultra-violet infrared mixing in theories with 
generic interaction of the Gross-Neveu type and allow consistent renormalization to all orders of perturbation. 

The free action becomes after vulcanization
\begin{equation}\label{sfree}
S_{free} = \int d^2 x \psib^a(x)\lbt\ps +\mu+\Omega\xts\rbt\psi^a  (x)
\end{equation}
where $\xt=2\Theta^{-1}x$ and $\Theta=\begin{pmatrix}\phantom{-}0&\theta\\-\theta&0\end{pmatrix}$ and
$a$ is a color index which takes values 1, ... $N$. The corresponding propagator $G$ is 
diagonal in this color index, so we omit it in this section\footnote{There
is no star product in these formulas, since we reduce quadratic expressions in the fields
in non-commutative theory to usual integrals with ordinary products.}.
To compute this propagator we write as in the
commutative case:
\begin{eqnarray}
G&=& \lbt\ps+\mu+\Omega\xts\rbt^{-1}=\lbt-\ps+\mu-\Omega\xts\rbt  . Q^{-1},
\nonumber\\
Q&=& \lbt\ps+\mu+\Omega\xts\rbt\lbt-\ps+\mu-\Omega\xts\rbt
  \nonumber\\
  &=&\mathds{1}_{2}\otimes\lbt
  p^{2}+\mu^{2}+\frac{4\Omega^{2}}{\theta^{2}}x^{2}\rbt
  +\frac{4\imath\Omega}{\theta}\gamma^{0}\gamma^{1}\otimes{\text{Id}}
  +\frac{4\Omega}{\theta}\mathds{1}_{2}\otimes  L_{2},
\end{eqnarray}
where $L_{2}=x^{0}p_{1}-x^{1}p_{0}$. 

To invert $Q$ we use again the Schwinger trick and obtain:
\begin{lemma}
\label{FermioXspace} We have:
\begin{eqnarray}
G(x,y) &=& -\frac{\Omega}{\theta\pi}\int_{0}^{\infty}\frac{dt}{\sinh(2\Ot
t)}\, e^{-\frac{\Ot}{2}\coth(2\Ot t)(x-y)^{2}+\imath\Ot
x\wedge y}
\\ 
&&    \lb\imath\Ot\coth(2\Ot t)(\xs-\ys)+\Omega(\xts-\yts)- \mu \rb
e^{-2\imath\Ot
t\gamma^{0}\gamma^{1}}e^{-t\mu^{2}}  \nonumber
\end{eqnarray}
It is also convenient to write $G$ in terms of commutators:
\begin{eqnarray}    
G(x,y)  &=&-\frac{\Omega}{\theta\pi}\int_{0}^{\infty}dt\,\lb \imath\Ot\coth(2\Ot
t)\lsb\xs, \Gamma^t  \rsb(x,y) \right.
\nonumber\\
&&
\left. +\Omega\lsb\slashed{\tilde{x}}, \Gamma^t \rsb(x,y)  -\mu \Gamma^t (x,y)  \rb
e^{-2\imath\Ot t\gamma^{0}\gamma^{1}}e^{-t\mu^{2}}, 
\label{xfullprop}
\end{eqnarray} 
where
\begin{eqnarray}
\Gamma^t (x,y)  &=&
\frac{1}{\sinh(2\Ot t)}\,
e^{-\frac{\Ot}{2}\coth(2\Ot t)(x-y)^{2}+\imath\Ot x\wedge y}
\end{eqnarray}
with $\Ot=\frac{2\Omega}{\theta}$ and $x\wedge y=x^{0}y^{1}-x^{1}y^{0}$.\\
\end{lemma}

\noindent
{\proof} 
The proof follows along the same lines than for Lemma \ref{HinXspace}
and is given in detail in Appendix \ref{app2}. Note that the constant term 
$e^{-2\imath\Ot t\gamma^{0}\gamma^{1}}$ is developped in (\ref{expgamma01}).
\qed
\medskip

\subsection{Bosonic Kernel in the Matrix Basis}

Let $H$ be as in Lemma \ref{HinXspace}, with $\Omega\rightarrow 
\frac{2\Omega}{\theta}$ and $B\rightarrow\frac{2B}{\theta}$. A 
straightforward computation shows that in the matrix basis we have:
\begin{align}
  H_{m,m+h;l+h,l}=&\frac{2}{\theta}
  (1+\Omega^2)(2m+h+1)\delta_{m,l}-\frac{4Bh}{\theta}~\delta_{m,l}\label{eq:HBini}
  \\
  &\hspace{-1cm}-\frac{2}{\theta}(1-\Omega^2)[\sqrt{(m+h+1)(m+1)}~\delta_{m+1,l}+
  \sqrt{(m+h)m}~\delta_{m-1,l}].\nonumber
  \label{HinMat}
\end{align}
Notice that in the limiting case $\Omega=B=1$ the operator becomes diagonal.

The corresponding propagator in the matrix basis for $B=0$ 
can be found in \cite{GrWu04-3} and \cite{GrWu03-2}. The result is
that the only non zero matrix elements of the 
exponential are:  
\begin{eqnarray} 
    &&[e^{-\frac{t\theta}{8\Omega}H}]_{m,m+h;l+h,l}=\\
    \nonumber 
    &&\sum^{\min(m,l)}_{u=\max(0,-h)}
    \Big{(}\frac{4\Omega}{(1+\Omega)^2}\Big{)}^{h+2u+1}
    \Big{(}\frac{1-\Omega}{1+\Omega}\Big{)}^{m+l-2u}
    {\cal E}(m,l,h,u){\cal A}(m,l,h,u)
\end{eqnarray}     
with
\begin{equation}\label{eq:A}
{\cal A}(m,l,h,u)=\frac{\sqrt{m!(m+h)!l!(l+h)!}}{(m-u)!(l-u)!(h+u)!u!},
\end{equation}
\begin{equation}
{\cal E}(m,l,h,u)=\frac{e^{-t(\frac{h+1}{2}+u)}(1-e^{-t})^{m+l-2u}}
{(1-(\frac{1-\Omega}{1+\Omega})^2)e^{-t})^{m+l+h+1}}.
\end{equation}

Having in mind the slicing of the propagators needed to 
carry out the renormalization (see \cite{Rivasseau2005bh} for the 
$\phi^4$ case), we will use a slightly different 
representation of the propagator:
\begin{equation}
  H^{-1}=\frac{\theta}{8\Omega}\int_{0}^{1}\frac{d\alpha}
  {1-\alpha}(1-\alpha)^{\frac{\theta}{8\Omega}H}.
\end{equation} 
One has then the lemma:
\begin{lemma}\label{lemma:B0bis}
  Let $H$ be given by equation (\ref{eq:HBini}) with $B=0$. We have:
  \begin{equation} 
    [(1-\alpha)^{\frac{\theta}{8\Omega}H}]_{m,m+h;l+h,l}
    =\sum_{u=max(0,-h)}^{min(m,l)}{\cal A}(m,l,h,u)
    \Big{(}C\frac{1+\Omega}{1-\Omega}\Big{)}^{m+l-2u}{\cal E}(m,l,h,u)
  \end{equation} 
  with ${\cal A}(m,l,h,u)$ as before, $C=\frac{(1-\Omega)^2}{4\Omega}$, and
  \begin{equation}
    {\cal E}(m,l,h,u)=\frac{(1-\alpha)^{\frac{h+2u+1}{2}}\alpha^{m+l-2u}}
    {(1+C\alpha)^{m+l+h+1}}.
  \end{equation}
\end{lemma}
The proof is given in Appendix \ref{app3} below. Extending to the  $B\ne 0$ case, we get
easily the following corollary, useful for studying the Gross-Neveu model:
\begin{corollary}
  Let $B\neq 0$. Denote $H_0=H|_{B=0}$ We have:
  \begin{equation}
    [(1-\alpha)^{\frac{\theta}{8\Omega}H}]_{m,m+h;l+h,l}=
    [(1-\alpha)^{\frac{\theta}{8\Omega}H_0}]_{m,m+h;l+h,l}
    (1-\alpha)^{-\frac{4B}{8\Omega}h}.
  \end{equation}
\end{corollary}

\subsection{Fermionic Kernel in the Matrix Basis}

Let $L_{2}=-\imath(x^{0}\partial_{1}-x^{1}\partial_{0})$. The inverse of the quadratic form
\begin{equation}
\Delta= Q- \frac{4 \imath \Omega} {\theta} \gamma^0\gamma^1= p^{2}+\mu^{2}+\frac{4\Omega^{2}}{\theta^2} x ^{2} +\frac{4B}{\theta}L_{2}
\end{equation}
is given by the previous section:
\begin{align}
  \label{eq:propinit}
  \Gamma_{m, m+h; l + h, l} 
  &= \frac{\theta}{8\Omega} \int_0^1 d\alpha\,  
  \dfrac{(1-\alpha)^{\frac{\mu^2 \theta}{8 \Omega}-\frac{1}{2}}}{  
    (1 + C\alpha )} 
  \Gamma^{\alpha}_{m, m+h; l + h, l}\;,
  \nonumber\\
  \Gamma^{(\alpha)}_{m, m+h; l + h, l}
  &= \left(\frac{\sqrt{1-\alpha}}{1+C \alpha} 
  \right)^{m+l+h}\left( 1-\alpha\right)^{-\frac{Bh}{2\Omega}} \\
  &
  \sum_{u=0}^{\min(m,l)} {\cal A}(m,l,h,u)\ 
  \left( \frac{C \alpha (1+\Omega)}{\sqrt{1-\alpha}\,(1-\Omega)} 
  \right)^{m+l-2u}\;,
  \label{eq:propinit-b}
\end{align}
where ${\cal A}(m,l,h,u)$ is given by (\ref{eq:A}) and $C$ is defined in Lemma
\ref{lemma:B0bis}.\\

The Fermionic propagator $G$ (\ref{xfullprop}) in matrix space
can be deduced from this kernel. One should simply take $B= \Omega$,
add the missing $\gamma^0 \gamma^1$ term, and compute the action of
$-\ps-\Omega\xts+\mu$ on $\Gamma$. Hence we have to compute $\lsb x^{\nu},\Gamma\rsb$ in the
matrix basis. It is easy to express the multiplicative operator $x^{\nu}$ in
this matrix basis. Its commutator with $\Gamma$ follows from
\begin{align}
  \lsb x^{0},\Gamma\rsb_{m,n;k,l}=&2\pi\theta\sqrt\frac{\theta}{8}\lb\sqrt{m+1}
  \Gamma_{m+1,n;k,l}-\sqrt{l}\Gamma_{m,n;k,l-1}+\sqrt{m}\Gamma_{m-1,n;k,l}
\right.\nonumber\\
&-\sqrt{l+1}\Gamma_{m,n;k,l+1}+\sqrt{n+1}\Gamma_{m,n+1;k,l}-\sqrt{k}
\Gamma_{m,n;k-1,l}\nonumber\\
&\left.+\sqrt{n}\Gamma_{m,n-1;k,l}-\sqrt{k+1}
  \Gamma_{m,n;k+1,l}\rb,\label{x0Gamma}\\
  \lsb
  x^{1},\Gamma\rsb_{m,n;k,l}=&2\imath\pi\theta\sqrt\frac{\theta}{8}\lb\sqrt{m+1}
  \Gamma_{m+1,n;k,l}-\sqrt{l}\Gamma_{m,n;k,l-1}-\sqrt{m}
  \Gamma_{m-1,n;k,l} \right.
\nonumber\\
&+\sqrt{l+1}\Gamma_{m,n;k,l+1}
-\sqrt{n+1}\Gamma_{m,n+1;k,l}+\sqrt{k}\Gamma_{m,n;k-1,l}
\nonumber\\
&\left.+\sqrt{n}\Gamma_{m,n-1;k,l}-\sqrt{k+1}\Gamma_{m,n;k+1,l}\rb.
  \label{x1Gamma}
\end{align}
This leads to the formula for $G$ in matrix space:
\begin{lemma}Let $G_{m,n;k,l}$ be the matrix basis kernel of the operator\\
$\lbt\ps+\Omega\xts+\mu\rbt^{-1}$. We have:
\begin{eqnarray}
G_{m,n;k,l}&=& 
-\frac{2\Omega}{\theta^{2}\pi^{2}} \int_{0}^{1} 
d\alpha\, G^{\alpha}_{m,n;k,l}  
\nonumber\\
G^{\alpha}_{m,n;k,l}&=&\lbt\imath\Ot\frac{2-\alpha}{\alpha}\lsb\xs,
\Gamma^{\alpha}\rsb_{m,n;k,l}
+\Omega\lsb\slashed{\tilde{x}},\Gamma^{\alpha}\rsb_{m,n;k,l} - \mu\,\Gamma^{\alpha}_{m,n;k,l}\rbt
\nonumber\\
&&\times\lbt\frac{2-\alpha}{2\sqrt{1-\alpha}}
\mathds{1}_{2}-\imath\frac{\alpha}{2\sqrt{1-\alpha}}\gamma^{0}\gamma^{1}
\rbt.\label{eq:matrixfullprop}
\end{eqnarray}
where $\Gamma^{\alpha}$ is given by (\ref{eq:propinit-b}) and the commutators 
by formulae (\ref{x0Gamma}) and (\ref{x1Gamma}).
\end{lemma}
The first two terms in (\ref{eq:matrixfullprop}) contain commutators and are grouped together
under the name $G^{\alpha, {\rm comm}}_{m,n;k,l}$. The last term is called 
$G^{\alpha, {\rm mass}}_{m,n;k,l}$.  Hence
\begin{eqnarray}\label{commterm}
G^{\alpha, {\rm comm}}_{m,n;k,l}&=& \lbt\imath\Ot\frac{2-\alpha}{\alpha}\lsb\xs,
\Gamma^{\alpha}\rsb_{m,n;k,l} +\Omega\lsb\slashed{\tilde{x}},\Gamma^{\alpha}\rsb_{m,n;k,l} \rbt  \nonumber\\
&&\times\lbt\frac{2-\alpha}{2\sqrt{1-\alpha}}
\mathds{1}_{2}-\imath\frac{\alpha}{2\sqrt{1-\alpha}}\gamma^{0}\gamma^{1} \rbt.
\end{eqnarray}
\begin{eqnarray}\label{massterm}
G^{\alpha, {\rm mass}}_{m,n;k,l}&=& - \mu\, \Gamma^{\alpha}_{m,n;k,l}
\times\lbt\frac{2-\alpha}{2\sqrt{1-\alpha}}
\mathds{1}_{2}-\imath\frac{\alpha}{2\sqrt{1-\alpha}}\gamma^{0}\gamma^{1} \rbt.
\end{eqnarray}

\section{Non-Commutative Gross-Neveu Model\\ in the Matrix Basis}
\setcounter{equation}{0}

The Euclidean two-dimensional Gross-Neveu model is written in terms of $N$ pairs of conjugate
Grasmmann fields $\psib^a , \psi^a$, $a=1, ... N$. In the commutative case 
the interaction has the following "$N$-vector" form:
\begin{equation}
\lambda\lbt \sum_a \psib^a  \psi^a \rbt^{2}(x).
\end{equation}

There are several non-commutative generalizations of this interaction, 
depending on how to put the star product with respect to color 
and conjugation. The most general action involving a vertex with two colors,
each present on one $\psib$ and one $\psi$ takes the form (using cyclicity of the integral trace
of star products):
\begin{align}\label{completeaction}
S&= S_{free} + \int \Tr   V\lbt\psib,\psi\rbt \ ,
\nonumber\\
 V\lbt\psib,\psi\rbt&=\sum_{a,b}\lambda_{1}\psib^a\star\psi^{a}\star\psib^b\star\psi^{b}
      +\lambda_{2}\psib^a\star\psi^{b}\star\psib^b\star\psi^{a}\nonumber\\
      &+\lambda_{3}\psib^a\star\psib^b\star\psi^{a}\star\psi^{b}
      +\lambda_{4}\psib^a\star\psib^b\star\psi^{b}\star\psi^{a}
\end{align}
where $S_{free}$ is defined in (\ref{sfree}).
Recall that in the matrix basis the Grasmmann fields 
$\psib^a , \psi^a$, $a=1, ... N$ are Grassmann matrices 
$\psib^a_{mn}, \psi^a_{mn}$, $a=1, ... N$, $m,n \in {\mathbb N}$.

\begin{figure}
\begin{center}
\includegraphics[width=6cm]{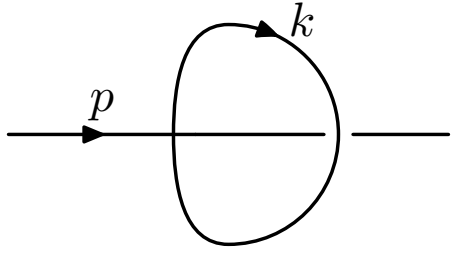}
\end{center}
\caption{The non-planar tadpole}
\label{figtadpole}
\end{figure}

In the cases $\lambda_3 =\lambda_4 = 0$, there are no non planar tadpole of the type of
figure \ref{figtadpole}
which lead to  infrared-ultraviolet mixing in $\phi^4_4$. 
Hence one may superficially conclude that there is no need
to vulcanize the free action, hence to use (\ref{sfree}). However even for
$\lambda_3 =\lambda_4 = 0$, four point functions lead
to "logarithmic" IR/UV  mixing,
hence to "renormalons" effects (large graphs with amplitudes of size $n!$ at order $n$).
Since we want anyway to renormalize generic Gross-Neveu actions in which 
$\lambda_3 \ne 0$ or $\lambda_4 \ne 0$, we shall always use the vulcanized free action
and propagator.

A proof of the BPHZ renormalization theorem according to the multiscale analysis \cite{Rivasseau2005bh}
decomposes into two main steps. First one has to prove bounds on the sliced propagator;
then using these bounds one has to prove that irrelevant operators in the multiscale analysis
give rise to convergent sums, and that this is also the case for marginal and relevant operators
{\it after} subtracting a singular part of the same form than the original action.

In this paper we provide the first part of this proof, namely the appropriate bounds.
The rest of the proof of renormalizability to all orders of the generic model (\ref{completeaction})
is postponed to a future paper \cite{RenNCGN05}.
However we illustrate on a particular example below why the bounds of 
this paper (which are optimal in a certain sense) should be sufficient for this task, 
which is however significantly more difficult
than the correpsonding task for $\phi^4_4$.

The multiscale  slice decomposition is performed as in \cite{Rivasseau2005bh}
\begin{equation}
  \label{eq:slices}
  \int_{0}^1 d\alpha = \sum_{i=1}^\infty \int_{M^{-i}}^{M^{-i+1}} 
  d\alpha
\end{equation}
and leads to the following propagator for the $i^{\text{th}}$ slice: 
\begin{align}
  \Gamma^i_{m,m+h,l+h,l} 
  &=\frac{\theta}{8\Omega}  \int_{M^{-i}}^{M^{-i+1}} d\alpha\; 
  \dfrac{(1-\alpha)^{\frac{\mu_0^2 \theta}{8 \Omega}-\frac{1}{2}}}{  
    (1 + C\alpha )} 
  \Gamma^{(\alpha)}_{m, m+h; l + h, l}\;.
  \label{prop-slice-i}
\end{align}
\begin{eqnarray}
G_{m,n;k,l}&=& \sum_{i=1}^\infty G^i_{m,n;k,l} \ ; \ G^i_{m,n;k,l} = 
-\frac{2\Omega}{\theta^{2}\pi^{2}} \int_{M^{-i}}^{M^{-i+1}} 
d\alpha\, G^{\alpha}_{m,n;k,l}  
\label{eq:matrixfullpropsliced}
\end{eqnarray}

We split $G$ according to (\ref{commterm}) and  (\ref{massterm}),
and we now bound $\vert G^i_{m,n;k,l}\vert$. We define $h= n-m$ and $p=l-m$.
By obvious symmetry of the integer indices we can assume $h \ge 0 $, and 
$p \ge 0$, so that   the smallest  of the four integers $m,n,k,l$ is $m$ and
the largest is $k=m+h+p$. The main result of this paper is the following bound:

\begin{theorem}\label{maintheorem}
Under the assumed conditions $h =n-m\ges 0 $ and $p=l-m \ges 0$
the Gross-Neveu propagator in a slice $i$ obeys the bound
\begin{eqnarray}\label{mainbound1}  
\vert G^{i,{\rm comm}}_{m,n;k,l}\vert&\les&   
K M^{-i/2} \bigg(  \frac{\exp \{- \frac{c p ^2  }{1+ kM^{-i}}
- \frac{ c M^{-i}}{1+k} (h - \frac{k}{1+C})^2  )  \}}{(1+\sqrt{ kM^{-i}}) }  
\nonumber\\
&&+ e^{- c k M^{-i} - c  p }\bigg).
\end{eqnarray}
for some (large) constant $K$ and (small) constant $c$ which depend only on $\Omega$.
Furthermore the part with the mass term has a slightly different bound:
\begin{equation} \label{mainbound2}  
\vert  G^{i,{\rm mass}}_{m,n;k,l}\vert\les   
K M^{-i} \bigg(  \frac{\exp \{- \frac{c p ^2  }{1+ kM^{-i}}
- \frac{ c M^{-i}}{1+k} (h - \frac{k}{1+C})^2  )  \}}{1+\sqrt{ kM^{-i}}}  
+ e^{- c k M^{-i} - c  p }\bigg).
\end{equation}
\end{theorem}

The rest of the paper is devoted to the proof of this theorem.
We give this proof only for $i\gg 1$, the ``first slices'' 
being unimportant for renormalization.

In the rest of this section we indicate
how this bound leads to efficient power counting estimates
for renormalization.

Recall first that for any non-commutative Feynman graph $G$ we can define
the genus of the graph, called $g$ and the number of faces ``broken by external legs", called $B$
as in \cite{GrWu04-3}-\cite{Rivasseau2005bh}.
We have $g \ge 0$ and $B\ge 1$. 
The power counting established for $\phi^4_4$ in 
\cite{GrWu04-3}-\cite{Rivasseau2005bh} 
involves the superficial degree of divergence of a graph
\begin{equation}
\omega (G) = (2 -N/2)  - 4 g  -2(B-1) \ ,
\end{equation}
and it is positive only for $N=2$ and $N=4$ subgraphs
with $g=0$ and $B=1$. These are the only non-vacuum graphs
that have to be renormalized. We expect the same conclusion for the two-dimensional
Gross-Neveu model, since this holds for the commutative counterpart of these models.

Let us sketch now why the multislice analysis based on bound (\ref{maintheorem}) proves that a 
graph  with internal propagators in slice $i >> 1$ and external legs in slice 1 
with $N\ge 6$ or $N=4$ and $g\ge 1$ does not require renormalization.
First remark that the second term in bound (\ref{maintheorem})  
gives exactly the same decay proved in \cite{Rivasseau2005bh}.
The $O(1)$ decay in $p$ means that the model is quasi-local in the sense
of \cite{GrWu03-2}. Hence all indices except those of independent faces cost $O(1)$ to sum.
Each main "face index" is summed with the scale decay $k M^{-i}$, hence each face sum costs $M^i$
in two dimensions. Combined with the $M^{-i/2}$ scaling factors of the propagators 
in (\ref{maintheorem}) one recovers the usual power counting in $\omega$.

\begin{figure}
  \begin{center}
    \includegraphics[scale=.7]{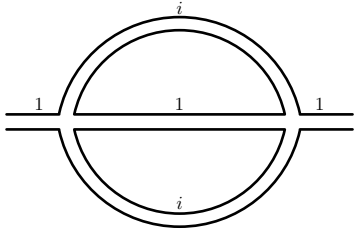}
  \end{center}
  \caption{The Sunset Graph}
  \label{figsunset}
\end{figure}

Hence let us concentrate on the more difficult case
of the first part of bound (\ref{maintheorem}), and  for instance 
consider the ``sunset" graph $G$ of Figure \ref{figsunset}

When the scales of the three internal lines are roughly identical, this graph should be
renormalized as a two-point subgraph and nothing particularly new
happens. But something new occurs when the two {\it exterior} lines have scales $i>>1$ and the interior line
has scale $1$ (like the external legs), as shown on the figure.
In the usual multiscale analysis we do not have a divergent two point 
subgraph in the traditional sense.The subgraph made of the two external lines
 is ``dangerous", i.e. has all internal scales above all external ones. But this graph 
has  $N=4$ and $g=0$, $B=2$, and it should {\it not} and in fact {\it cannot } be renormalized.
However applying bound (\ref{mainbound1})  the sum over $i$ diverges logarithmically! 
Indeed the sums over the $p$
indices cost only $O(1)$ as usual for a quasi-matrix model, 
but the two internal faces sums, together with their lines prefactor give
\begin{equation}
\sum_{k, \delta h = 0}^{\infty} 
  M^{-i/2} M^{-i/2}   e^{-M^{-i} k} \frac{e^{ -  (\delta h )^2 /k }}{1 + \sqrt k} = O(1) .
\end{equation}

What is the solution to this riddle?
In this case it is the full {\it two-point } subgraph $G$
which has to be renormalized. This works because the renormalization improvement 
brings modified ``composite propagators"  {\it solely}
on the exterior face of the graph \cite{GrWu04-3}. 
These improved propagators have scale $i$, hence they bring a factor $M^{-i}$, 
and the sum over $i$ converges.

One has to generalize this argument, and show that all the counterterms are of the right form
to complete the BPHZ theorem for this kind of models \cite{RenNCGN05}.
In short all dangerous subgraphs with $N=2,4$, $g=0$, $B=1$,  {\it and} $N=4$, $g=0$, $B=2$
should be renormalized, the last ones being renormalized by the corresponding 
{\it two-point} function counterterms.
This subtlety makes the multislice formulation of renormalization in the non-commutative 
Gross-Neveu model more complcated  than in the non-commutative $\phi^4_4$.

\section{Proof of Theorem \ref{maintheorem}}
\setcounter{equation}{0}

We cast the propagator for $B=\Omega$ in the following form:
\begin{equation}
  \Gamma=\int_{0}^{1}~d\alpha\frac{(1-\alpha)^{-1/2}}{1+C\alpha}\Gamma^{\alpha}
\end{equation}
with:
\begin{equation}\label{expressgamma}
  \Gamma^{\alpha}=\Big{(}\frac{\sqrt{1-\alpha}}{1+C\alpha}\Big{)}^{2m+p}
  \frac{1}{(1+C\alpha)^h}\sum_{u=o}^{m}
  \Big{(}\frac{\alpha\sqrt{C(1+C)}}
  {\sqrt{(1-\alpha)}}\Big{)}^{2m+p-2u}{\cal A}(m,m+p,h,u).
\end{equation}
We have:
\begin{equation}
\Gamma^{\alpha}=e^{(2m+p)\ln\frac{\sqrt{1-\alpha}}{1+C\alpha}-
h\ln(1+C\alpha)}\sum_{0\les v=m-u\les m}e^{(2v+p)\ln
\frac{\alpha\sqrt{C(1+C)}}{\sqrt{1-\alpha}}}{\cal A}(m,m+p,h,u).
\end{equation}
We consider the regime $\alpha \ll C \ll 1$ hence we limit ourselves 
as usually to a parameter $\Omega$ close to 1. We use Stirling's formula to write:
\begin{equation}
  \Gamma^{\alpha}\les K
  \sum_{0\les v\les k - h - p }e^{f(m,l,v)}\frac{\Big{(}
    (2\pi)^4  k    (k-h) (k-p) (k - h -p ) 
    \Big{)}^{1/4}}
  {\sqrt{(2\pi)^4v(v+p)(k-v-h -p )(k-v - p)}}
\end{equation}
with:
\begin{eqnarray}
  f&=&(2k -2h - p)\ln\frac{\sqrt{1-\alpha}}{1+C\alpha}-h\ln(1+C\alpha)+
  (2v+p)\ln\frac{\alpha\sqrt{C(1+C)}}{\sqrt{1-\alpha}}
  \nonumber\\
  &&+\frac{k - h -p }{2}\ln (k - h -p )
  +\frac{k-h}{2}\ln(k-h)+\frac{k- p }{2}
  \ln(k - p )
  \nonumber\\
  &&+\frac{k}{2}\ln k
  -v\ln v-(v+p)\ln(v+p) - (k-v- p )\ln(k-v - p )
  \nonumber\\
  &&-(k-v-h - p )\ln(k-v-h - p ).
\end{eqnarray}
We define the reduced variables $x = v/k$,  $y =h/k$, 
$z=p/k$. These parameters live in the compact
simplex $0 \les x, y, z \les 1$,  $0 \les x+ y+ z \les 1$. In
our propagator bound we can replace the Riemann sum over $v$ with the 
integral. Since we have a bounded function on a compact interval, 
this is a rigorous upper bound (up to some inessential overall constant)
for $k$ large, which is the case of interest:
\begin{equation}\label{integralinx}
  \Gamma^{\alpha}\les \int_0^{1-y-z} dx  \frac{[(1-y)(1-z)(1-z-y)]^{1/4}}
  {[x(x+z)(1-x-z) (1-x-y-z)]^{1/2}} e^{k g(x,y,z)}
\end{equation}
with the function $g$ defined by:
\begin{eqnarray}
  g&=&(2-2y-z)\ln\frac{\sqrt{1-\alpha}}{1+C\alpha}
  +(2x+z)\ln\frac{\alpha\sqrt{C(1+C)}}{\sqrt{1-\alpha}}- y \ln(1+C\alpha)
  \nonumber\\
  &&+ \frac{1-y}{2}\ln(1-y) +\frac{1-z}{2}\ln(1-z)  +\frac{1-y-z}{2}\ln(1-y-z)
  \nonumber\\
  &&-x\ln x-(x+z)\ln(x+z)  -(1-x-z)\ln(1-x-z) 
  \nonumber\\
  &&-(1-x-y-z)\ln(1-x-y-z).
\end{eqnarray}
The differential is 
\begin{eqnarray}\label{diferential}
  dg &=& dx \left\{  \ln\frac{\alpha^2 C(1+C) (1-x-z) (1-x-y-z) }
    {(1-\alpha)x(x+z)}  \right\}\\
  &&+ dy \left\{  \ln\frac{(1+C\alpha)(1-x-y-z)}{(1-\alpha)\sqrt{(1-y)(1-y-z)}} 
  \right\}
\nonumber\\
&&+ dz \left\{  \ln\frac{\alpha\sqrt{C(1+C)}(1+C\alpha)(1-x-z)(1-x-y-z)}{(1-\alpha)(x+z)\sqrt{(1-z)(1-y-z)}} 
\right\}\nonumber.
\end{eqnarray}
The second derivatives are
\begin{eqnarray}\label{seconderivatives}
  \frac{\partial ^2g}{\partial x^2}&=&-\frac{1}{x}-\frac{1}{x+z}-\frac{1}{1-x-z}
  -\frac{1}{1-x-y-z},\nonumber\\
  \frac{\partial^2g}{\partial y^2}&=&\frac{1}{2(1-y)}+ \frac{1}{2(1-y-z)}- \frac{1}{1-x-y-z},\nonumber\\
  \frac{\partial^2g}{\partial z^2}&=&\frac{1}{2(1-z)}+ \frac{1}{2(1-y-z)}- \frac{1}{x+z} 
  - \frac{1}{1-x-z} - \frac{1}{1-x-y-z},\nonumber\\
  \frac{\partial^2g}{\partial x \partial y }&=&-\frac{1}{1-x-y-z},\nonumber\\
  \frac{\partial^2g}{\partial x \partial z }&=&-\frac{1}{x+z}-\frac{1}{1-x-z}-\frac{1}{1-x-y-z},\nonumber\\
  \frac{\partial^2g}{\partial y \partial z}&=&\frac{1}{2(1-y-z)}-\frac{1}{1-x-y-z}. 
\end{eqnarray}

\begin{lemma}
The function $g$ is concave in the simplex.
\end{lemma}
\prf
We have to prove that the quadratic form $-Q$
defined by the 3 by 3 symmetric matrix of the second derivatives is negative in the whole simplex.
In others words we should prove that the oppposite quadratic form
\begin{eqnarray}
  \label{concave}
  Q&=&[\frac{1}{x} +\frac{1}{x+z}+  \frac{1}{1-x-z}+\frac{1}{1-x-y-z}]u^2
  \nonumber\\
  &+& [\frac{1}{1-x-y-z} - \frac{1}{2(1-y)} - \frac{1}{2(1-y-z)}] v^2
  \nonumber\\
  &+& [ \frac{1}{x+z} 
  +\frac{1}{1-x-z} +\frac{1}{1-x-y-z} - \frac{1}{2(1-z)}- \frac{1}{2(1-y-z)} ] w^2
  \nonumber\\
  &+& \frac{2}{1-x-y-z} uv + [ \frac{2}{x+z} + \frac{2}{1-x-z} + \frac{2}{1-x-y-z} ] uw
  \nonumber\\
  &+& [\frac{2}{1-x-y-z}-\frac{1}{1-y-z}] vw 
\end{eqnarray}
is positive. But we have
\begin{eqnarray}
  \label{concave1}
  Q&=& \frac{(u+v+w)^2}{1-x-y-z} + \frac{(u+w)^2}{(x+z)(1-x-z)}  +
  \frac{u^2}{x}
  - \frac{v^2}{2(1-y)}  - \frac{(v+w)^2}{2(1-y-z)}
  \nonumber\\
  &\ge&
  \frac{(u+v+w)^2}{1-x-y-z} + \frac{(u+w)^2}{2(x+z)} +
  \frac{u^2}{2 x} - \frac{v^2}{2(1-y)} - \frac{(v+w)^2}{2(1-y-z)}
  \nonumber\\
  &=& \frac{1}{2}\left( \left[ \frac{(u+v+w)^2}{1-x-y-z} - \frac{v^2}{1-y} + \frac{(u+w)^2}{x+z}\right] \right.
  \nonumber\\&& \quad \quad \quad + \left. \left[ \frac{(u+v+w)^2}{1-x-y-z} - \frac{(v+w)^2}{1-y-z} + \frac{u^2}{x}\right]\right)
  \nonumber\\
  &=& \frac{1}{2}\left(\frac{1}{1-y}\left[\sqrt{\frac{x+z}{1-x-y-z}}(u+v+w) + \sqrt{\frac{1-x-y-z}{x+z}}(u+w)\right]^2\right.
  \nonumber\\
  &+& \frac{1}{1-y-z}\left. \left[ \sqrt{\frac{x}{1-x-y-z}}(u+v+w) + \sqrt{\frac{1-x-y-z}{x}}u\right]^2\  \right)
  \nonumber\\
  &\ge& 0.
\end{eqnarray}
\hfill\qed

\begin{lemma}\label{uniquemaximum}
  The only critical point of the function in the closed simplex is at $x_0 = \frac{C\alpha}{1 + C\alpha}$, $y_0= \frac{1}{1 + C}$, $z=0$,
  where the function $g=0$.
\end{lemma} 
\prf 
One can easily check that:
\begin{equation}
  g(x_0,y_0,0)= \partial_x g(x_0,y_0,0)= \partial_{y}g(x_0,y_0,0)=
  \partial_{z}g(x_0,y_0,0)=0.
\end{equation}
The unicity follows from the concavity of the function $g$.
\hfill\qed
\bigskip\\
Our bound on $g$ will be inspired by the steepest descent method around the critical point.
We divide now the simplex into:
\begin{itemize}
\item the neighborhood of the maximum. We call this region the ``mountain top''.
  It corresponds to $\delta x=\vert x-x_0 \vert\ll\alpha $, $\delta y=\vert
  y-y_0\vert\ll O(1)$, $z\ll\alpha$.
  For aesthetic reasons we prefer to use a reference quadratic form $Q_0$
  to define a smooth border of this region. Hence putting $X=(\delta x, \delta y , z)$ we
  define the mountain top by the condition
  \begin{equation}\label{defmountaintop}
    XQ_0^{\phantom{0}t}\!X  = \frac{(\delta x)^2  + z^2}{\alpha^2} + (\delta y)^2 \les \eta
  \end{equation}
  where $\eta$ is a small constant,

\item the rest of the simplex. This region is defined by $XQ_0^{\phantom{O}t}\!X\ge \eta $. 
\end{itemize}

\subsection{The ``Mountain Top''}
In this region we use the Hessian approximation and check that the cubic
correction terms are small with respect to this leading order.

\begin{lemma} In the mountain top region, for some small constant $c$ 
  (which may depend on $C$, hence on $\Omega$):
  \begin{equation}\label{mainhessianbound}
    g(X) \le  - c \alpha XQ_0^{\phantom{O}t}\!X  =  - c \alpha [\frac{(\delta x)^2  + z^2}{\alpha^2} + (\delta y)^2 ]
  \end{equation}
\end{lemma}

\prf
>From (\ref{seconder}) we evaluate the second order derivatives of $g$ at leading order 
in $\alpha$ at the maximum: 
\begin{align}
  \frac{\partial ^2g}{\partial x^2} &\approx -\frac{2}{\alpha C
  };&\frac{\partial^2g}{\partial x \partial z }&\approx -\frac{1}{\alpha
    C},\nonumber\\
  \frac{\partial^2g}{\partial z^2} &\approx -\frac{1}{\alpha C};& \frac{\partial^2g}{\partial x \partial y } &\approx
  -\frac{1+C}{C},\nonumber\\  
  \frac{\partial^2g}{\partial y^2} &\approx -\alpha\frac{1+C}{C};&  
  \frac{\partial^2g}{\partial y \partial z}&\approx -\frac{1+C}{2C}.\label{seconder}
\end{align}

It is easy to diagonalize
the corresponding 3 by 3 quadratic form, and to check that
in the neighborhood of the maximum it is smaller than $- c \alpha  Q_0$ 
for some small constant $c$:
\begin{equation} 
  \label{hessianbound}
  g_{Hessian}(\delta x,\delta y,z)= X  Q_{Hessian}^{\phantom{Hessian}t}\!X  
  \les -\alpha c XQ_0^{\phantom{O}t}\!X  =-\alpha c
  \Big{(} \frac{(\delta x)^2 + z^2}{\alpha^2}
  +(\delta y)^2 \Big{)},
\end{equation}
where $g_{Hessian}$ is the Hessian approximation to the function $g$.
It is easy to check from the expression (\ref{diferential}) of the differential $dg$
that the third order derivatives scale in the appropriate way so that choosing 
the constant $\eta$ small enough in (\ref{defmountaintop}), 
the function $g$ obeys the same bound than (\ref{hessianbound}) with 
a slightly different constant $c$. 
\hfill\qed

\subsection{The Rest of the Simplex}

To bound the function $g$ in the whole simplex we use the previous notation $X=(\delta x, \delta y, z)$. 
Drawing the segment from point $X$ to the origin (i.e. the mountain top), we
cross the border of the mountain top at $X_0= \lambda X$ with $X_{0}Q_0^{\phantom{O}t}\!X_{0} =\eta$. We
define $X_1= X_0/2= (\lambda/2) X $.  $X$ out of the mountain top means that
$\lambda = \sqrt{\eta/XQ_0^{\phantom{O}t}\!X}\les 1$. 

\begin{lemma}\label{firstoutbound}
Out of the mountain top region the function $g(X) = g(\delta x, \delta y, z)$ 
obeys the bound, for some small enough constant $c$:
\begin{equation}\label{goodlargefieldbound}
g(\delta x, \delta y, z) \les - c (\alpha   + \vert \delta x \vert + z).
\end{equation}
\end{lemma}
\prf
We use concavity of the function on the segment considered, which means that the function $g$ 
is below its first order Taylor approximation at $X_1$:
\begin{equation} 
  g(X) \les g(X_1) +  \langle dg(X_1),X-X_1\rangle. 
\end{equation}
At $X_1$ the Hessian approximation of $g$ is valid, say up to a factor 2. Hence
\begin{equation} 
  g(X) \les (1/2)[ X_1 Q_{Hessian}^{\phantom{Hessian}t}\!X_1  +  2(X-X_1) Q_{Hessian}^{\phantom{Hessian}t}\!X_1 ].
\end{equation} 
Using (\ref{hessianbound}) we can relate $Q_{Hessian}$ to our reference quadratic form 
$Q_0$ up to a constant and get for some small $c$:
\begin{eqnarray} 
  g(X) &\les& - c \alpha [ X_1 Q_{0}^{\phantom{0}t}\!X_1  +  2(X-X_1) Q_{0}^{\phantom{0}t}\!X_1 ] 
  \nonumber\\ &=& - c \alpha  [(\eta/4)  + \lambda(1-\lambda/2)XQ_0^{\phantom{0}t}\!X ] 
  \nonumber\\ &=& - c' \alpha  [1  + \sqrt{XQ_0^{\phantom{0}t}\!X}] 
\end{eqnarray} 
for some constant $c'$ smaller than $c$. In the last line we used $1-\lambda/2\ges 1/2$ and
$\lambda = \sqrt{\eta/XQ_0^{\phantom{0}t}\!X}$.
Finally 
\begin{equation}
  \alpha \sqrt{X Q_0 ^tX} \ge  \sqrt{(\delta x)^2  + z^2}\ges (\vert \delta x \vert  + z)/\sqrt{2}
\end{equation}
completes the proof of (\ref{goodlargefieldbound}).
\hfill\qed

\subsection{Integration on $\boldsymbol{x}$}
It remains now to prove some explicit decay of the function
$G$ in the variable $z$ after integration in $x$ in (\ref{integralinx}). 
The decay in $z$ is necessary to prove that the model is a quasi-local matrix model 
in the sense of \cite{GrWu03-1}. For the mountain top region,
we do not have any decay in $k$ so we want also to exhibit the decay in y.
\begin{lemma}\label{goodbound}
  For some large constant  $K$ and small constant $c$, under the condition $\alpha k \ge 1$
  we have
  \begin{equation}\label{auxbound2} 
    \Gamma^{\alpha}\les  K \bigg(  \frac{\exp \{- \frac{c}{\alpha k} p^2  
      - \frac{ c \alpha}{k} (h - \frac{k}{1+C})^2  ) \}}{\sqrt{\alpha k}}  
    + e^{- c \alpha k - c  p }   \bigg).
  \end{equation}
\end{lemma}
\prf
In the integration on $x$ in (\ref{integralinx}) we can insert 
$1 = \chi + (1-\chi)$ where $\chi$ is the characteristic function of the mountain top.
In the first term we apply the bound (\ref{mainhessianbound}) and in the
second the bound (\ref{goodlargefieldbound}). In this second case 
we use a better estimation of the prefactors in front of $e^{kg}$ in
(\ref{integralinx}). Actually their expression in (\ref{integralinx}) leads to
a spurious logarithmic divergence due to the bad behaviour of the Stirling
approximation close to 0. We will use
\begin{eqnarray}
  K\sqrt{n+1}\,n^{n}e^{-n}\les&n!&\les K'\sqrt n\,n^{n}e^{-n},
\end{eqnarray}
wich leads to an integral of the type 
\begin{align}
  \int_{0}^{1}\frac{dx}{x+1/k}e^{-ckx}&\les
  K\int_{1/k}^{\infty}\frac{dx}{x}e^{-ckx},
\end{align}
which is bounded by a constant. Scaling back to the original variables
completes the proof of Lemma \ref{goodbound}.
\hfill\qed

\subsection{The Region $\boldsymbol{\alpha k\les 1}$}
In this region we do not need Stirling's formula at all.
It is easier to derive a direct simple bound on $\Gamma^{\alpha}_{m,m+h;k,m+p}$ (recall that $k=m+h+p$):
\begin{lemma}\label{lemma:boundakpetit}
  For $M$ large enough and $\alpha k< 1$, there exists constants $K$ and $c$  such that
  \begin{equation}
    \label{eq:propppetit}
    \Gamma^{\alpha}_{m,m+h;k,m+p}\les Ke^{-c(\alpha k+p)}.
  \end{equation}
\end{lemma}
{\proof} We assume $h\ges 0$ and $p\ges 0$. For $\alpha k\les 1$, 
the following crude bound on $\Gamma^{\alpha}$ follows easily from (\ref{expressgamma}):
\begin{equation}\label{eq:kpetit1}
  \Gamma^{\alpha}_{m,m+h;k, m+p}\les 
  e^{-\alpha(C+1/2)(2m+p)-(C/2)\alpha h}
  \sum_{u=0}^{m}\frac{X^{m-u}}{(m-u)!}\frac{Y^{m+p-u}}{(m+p-u)!}
\end{equation}
where $X=\frac{\alpha\sqrt{C(C+1)}}{\sqrt{1-\alpha}}\sqrt{m(m+h)}$ and $Y=\frac{\alpha\sqrt{C(C+1)}}{\sqrt{1-\alpha}}\sqrt{(m+p)(m+p+h)}$. 
\begin{align}
  \label{eq:kpetit2}
  \Gamma^{\alpha}_{m,m+h,l+h,l}&\les e^{-\alpha(C+1/2)(2m+p)-(C/2)\alpha h}
  \frac{\big( \frac{\alpha k \sqrt{C(C+1)}}{\sqrt{1-\alpha}} \big)^{p}}{p!}
  \sum_{u=0}^{m}\frac{\big( \frac{\alpha k \sqrt{C(C+1)}}{\sqrt{1-\alpha}} \big)^{2(m-u)}}{(m-u)!^{2}}
\end{align}
The sum over $u$ is bounded by a constant. For $C$ small (i.e. $\Omega $ close to 1) we have certainly
$\frac{\sqrt{C(C+1)}}{\sqrt{1-\alpha}}\les 1/2$ hence we get the
desired result.\qed
\bigskip\\
Combining with (\ref{auxbound2}) we conclude that Lemma \ref{goodbound} always holds:
\begin{lemma}
  \label{goodbound1}
  For some large constant  $K$ and small constant $c$
  we have
  \begin{equation}\label{auxbound3} 
    \Gamma^{\alpha}\les  K \bigg(  \frac{\exp \{- \frac{c}{1+\alpha k} p^2  
      - \frac{ c \alpha}{1+ k} (h - \frac{k}{1+C})^2  ) \}}{1 + \sqrt{\alpha k}}  
    + e^{- c \alpha k - c  p }\bigg).
  \end{equation}
\end{lemma}

\subsection{Numerator Terms}

In this section we check that the numerators in the Gross-Neveu propagator
bring the missing power counting factors, hence we complete the proof of
Theorem \ref{maintheorem}.

The bound (\ref{mainbound2}) is nothing but the direct consequence of
multiplying the bound of Lemma \ref{goodbound1} by the width $M^{-i}$
of the integration interval over $\alpha$. Hence we now prove (\ref{mainbound1}).

In the commutator terms (\ref{commterm})
the $\Omega$ and the $\imath\frac{\alpha}
{2\sqrt{1-\alpha}}\gamma^{0}\gamma^{1}$
are smaller by at least an $\alpha$ factor. Therefore the largest piece
is the $O(1/\alpha)[\tilde x , \Gamma]$ term. The $1/alpha$ factor 
compensates the  width $M^{-i}$
of the integration interval over $\alpha$. Hence we need to prove that a $[\tilde x , \Gamma]$
numerator adds  a factor $\sqrt{\alpha}$ to the bound of Lemma \ref{goodbound1}:

(The $\Omega$ and the $\imath\frac{\alpha}{2\sqrt{1-\alpha}}\gamma^{0}\gamma^{1}$
terms in (\ref{commterm}) have an additional factor $\alpha$, hence are much easier 
to bound and left to the reader. The bound for the mass term $\mu \Gamma$ just
involves multpliying the bound of Lemma \ref{goodbound1} by the appropriate weight
$M^{-i}$ coming from the $\alpha$ integration, which leads easily to (\ref{mainbound2}).

The commutator  $[\slashed x , \Gamma]$ involves terms like  
\begin{align}
&\sqrt{m+1}\Gamma_{m+1,n;k,l}-\sqrt{l}\Gamma_{m,n;k,l-1}\nonumber\\
=&\lbt\sqrt{m+1}-   \sqrt{l}\rbt\Gamma_{m,n;k,l-1} 
+ \sqrt{m+1}\lbt\Gamma_{m+1,n;k,l}-\Gamma_{m,n;k,l-1}\rbt. \label{eq:typnum}
\end{align}

\subsubsection{The first term}
The first  term is the easiest to bound. It is zero unless $p=l-m-1 \ges 1$.
In this case,  we have 
\begin{equation}
\sqrt{m+1}-   \sqrt{l} \les  \frac{2p}{1 + \sqrt l}.
\end{equation}

Using Lemma \ref{lemma:dl}: 
\begin{itemize}
\item On the mountain top we have $l =k-h \simeq \frac{C}{1+C} k$,
hence $\frac{2p}{1 + \sqrt l} \les O(1) \frac{p}{1 + \sqrt k}  $. An additional
factor $\alpha$ comes from (\ref{inboundd}), hence we have a bound in $O(1) \frac{\alpha  p}{1 + \sqrt k}$.
Using a fraction of the decay $ e^{-c \alpha p^2/k}$ from Lemma \ref{goodbound1} bouinds
$O(1) \frac{\alpha  p}{1 + \sqrt k}$ by $\sqrt{\alpha}$.

\item Out of the mountain top we have a factor $\alpha \sqrt{kl}$ which comes from 
(\ref{outboundd}), and we can use $p e^{-cp} \les  e^{-c'p}$ and 
$\sqrt{\alpha k} e^{-c\alpha k} \les  e^{-c'\alpha k}$. Hence the desired factor $\sqrt\alpha$
is obtained.

\end{itemize}

It remains to prove
\begin{lemma}\label{lemma:dl}
Let $(m,l,h)\in\mathds{N}^{3}$ with $p=l-m-1 \ges 1$. We have 
\begin{itemize}
\item On the mountain top
\begin{equation}\label{inboundd}
      \Gamma_{m,m+h;k,m+p}\les K\alpha\,\Gamma_{m,m+h;m+p-1+h,m+p-1},
    \end{equation}
  \item if $\Gamma$ is out the mountain top region
    \begin{equation}\label{outboundd}
      \Gamma_{m,m+h;k,m+p}\les K\alpha \sqrt{kl}\,\Gamma_{m,m+h;m+p-1+h,m+p-1}.  
    \end{equation}
  \end{itemize}
\end{lemma}
{\proof} Let $p\ges 1$. The kernel $\Gamma_{m,m+h;m+p+h,m+p}$ is given by
\begin{align}
  \Gamma_{m,m+h;m+p+h,m+p}=&\lbt\frac{\sqrt{1-\alpha}}{1+C\alpha}
  \rbt^{2m+p}(1+C\alpha)^{-h}\nonumber\\
  &\sum_{u=0}^{m}(\alpha D)^{2(m-u)+p}{\cal A}(m,m+p,h, u)  
\end{align}
with ${\cal
  A}$ defined by (\ref{eq:A}) and $D(\alpha)=\sqrt{\frac{C(C+1)}{1-\alpha}}$.
The scaling factor $(\alpha D)^{2(m-u)+p}$ reaches its maximum at
$u=m$. There we have a factor $(\alpha D)^{p}$. For $p\ges 1$,
we factorize $\alpha D$ to get
\begin{align}
  \Gamma_{m,m+h;m+p+h,m+p}=&\alpha
  D\lbt\frac{\sqrt{1-\alpha}}{1+C\alpha}\rbt^{2m+p}(1+C\alpha)^{-h}\nonumber\\
  &\sum_{u=0}^{m}(\alpha D)^{2(m-u)+p-1}{\cal A}(m,m+p,h, u).
\end{align}
Using $\binom{m+p}{m+p-u}=\frac{m+p}{m+p-u}\binom{m+p-1}{m+p-1-u}$ and $\binom{m+p+h}{m+p-u}=\frac{m+p+h}{m+p-u}\binom{m+p-1+h}{m+p-1-u}$, we have
\begin{align}
  \Gamma_{m,m+h;m+p+h,m+p}&\les K\alpha\lbt\frac{\sqrt{1-\alpha}}{1+C\alpha}\rbt^{2m+p-1}
  (1+C\alpha)^{-h}\sum_{u=0}^{m}(\alpha
  D)^{2(m-u)+p-1}\nonumber\\
  &\hspace{1cm}\times{\cal A}(m,m+p-1,h, u)\frac{\sqrt{(m+p)(m+p+h)}}{m+p-u}.
\end{align}
On the ``mountain top'' we express $m+p-u$ in term of $k$ and get
\begin{align}
  \frac{1}{m+p-u}&=\frac{1}{k-u}\ \frac{1}{1-\frac{h}{k-u}}.
\end{align}
Moreover $u\simeq\alpha k$ and $(k-u)^{-1}\simeq
k^{-1}$, which leads to  (\ref{inboundd}).

Out of the mountain top we use simply $m+p-u \ges 1$ and\\$\sqrt{(m+p)(m+p+h)}= \sqrt{kl}$,
and (\ref{outboundd}) follows.
\qed

\subsubsection{The second term}

We now focus on the terms involving differences of $\Gamma$'s. For this we need
some identities on the combinatorial factor ${\cal A}$: 
\begin{align}
  {\cal A}(m,l,h,u)&=\frac{\sqrt{ml}}{u}{\cal A}(m-1,l-1,h+1,u-1),\text{ for
    $u\ges 1$},\label{eq:A-1}\\
  {\cal A}(m,l,h,u)&=\frac{\sqrt{m(m+h)}}{m-u}{\cal A}(m-1,l,h,u),\label{eq:Am-1}\\
  {\cal A}(m,l,h,u)&=\frac{\sqrt{m(m+h)l(l+h)}}{(m-u)(l-u)}
  {\cal A}(m-1,l-1,h,u).\label{eq:Aml-1}.
\end{align}
Let us recall $h=n-(m+1)$ and $p=l-(m+1)$. Then
$\Gamma_{m+1,n;k,l}=\Gamma_{m+1,m+1+h;l+h,l}$ and
$\Gamma_{m,n;k,l-1}=\Gamma_{m,m+h+1;l+h,l-1}$. 

Using Lemma (\ref{thm:numerator-terms}) we get:
\begin{equation}
  \sqrt{m+1}\lbt\Gamma_{m+1,n;k,l}-\Gamma_{m,n;k,l-1}\rbt
        \les K\sqrt{\alpha}\Gamma^{\alpha},
\end{equation}
  on the mountain top by (\ref{numtermmountop}) and the sequel, as well as 
out of the critical region by (\ref{numtermouttop}). The factors thus 
obtained together with Lemma (\ref{goodbound1})
complete the proof of (\ref{maintheorem}).

\begin{lemma}\label{thm:numerator-terms}
For $M$ large enough and $\Omega$ close enough to 1, there exist
$K$ such that 
\begin{equation}
\label{eq:fullprop-bound}
    G^{\alpha}_{m,n;k,l}\les K\sqrt{\alpha}\Gamma^{\alpha}_{m-1,n; k-1, l-2}.
  \end{equation}
\end{lemma}
{\proof} It remains to bound the second term in (\ref{eq:typnum}),\\ 
namely $\sqrt{m+1}\lbt\Gamma_{m+1,n;k,l}-\Gamma_{m,n;k,l-1}\rbt$. 
\begin{align}
  &\Gamma_{m+1,m+1+h;l+h,l}-\Gamma_{m,m+h+1;l+h,l-1}\\
  =&\lbt\frac{\sqrt{1-\alpha}}{1+C\alpha}\rbt^{m+1+l}(1+C\alpha)^{-h}
  \sum_{u_{1}=0}^{m+1}(\alpha D)^{m+1+l-2u_{1}}{\cal A}(u_{1};m+1,l,h)\nonumber\\
  &-\lbt\frac{\sqrt{1-\alpha}}{1+C\alpha}\rbt^{m+l-1}(1+C\alpha)^{-h-1}
  \sum_{u_{2}=0}^{m}(\alpha
  D)^{m+l-1-2u_{2}}{\cal A}(u_{2};m,l,h+1)\nonumber.
\end{align}

Thus we conclude, again up to boundary terms (treated in Subsection \ref{subsubboundary} below)
\begin{align}\label{boundaryterm1}
  &\Gamma_{m+1,m+1+h;l+h,l}-\Gamma_{m,m+h+1;l+h,l-1}\\
  &\les \lbt\frac{\sqrt{1-\alpha}}{1+C\alpha}\rbt^{m+l+1}(1+C\alpha)^{-h}
  \sum_{u_{1}=1}^{m+1}(\alpha D)^{m+1+l-2u_{1}}{\cal A}(u_{1};m+1,l,h)\nonumber\\
  &-\lbt\frac{\sqrt{1-\alpha}}{1+C\alpha}\rbt^{m+l-1}(1+C\alpha)^{-h-1}
  \sum_{u_{2}=0}^{m}(\alpha
  D)^{m+l-1-2u_{2}}{\cal A}(u_{2};m,l-1,h+1)\nonumber
\end{align}
Let $u_{1}=u+1$. Thanks to (\ref{eq:A-1}) we write the sum over $u_{1}$
as 
\begin{align}
  &\sum_{u_{1}=1}^{m+1}(\alpha D)^{m+1+l-2u_{1}}{\cal
    A}(u_{1};m+1,l,h)\nonumber\\
  =&\sum_{u=0}^{m}(\alpha D)^{m+l-1-2u}{\cal A}(u;m,l-1,h+1)\frac{\sqrt{l(m+1)}}{u+1}.
\end{align}
The difference in (\ref{boundaryterm1}) is now written as 
\begin{align}
  {\cal D}\defi&\lbt\frac{\sqrt{1-\alpha}}{1+C\alpha}\rbt^{m+l-1}(1+C\alpha)^{-h-1}
  \sum_{u=0}^{m}(\alpha
  D)^{m+l-1-2u}{\cal A}(u;m,l-1,h+1)\nonumber\\
  &\times\lb\lbt\frac{\sqrt{1-\alpha}}{1+C\alpha}\rbt^{2}(1+C\alpha)\frac{\sqrt{l(m+1)}}
  {u+1}-1\rb. \label{bondaryterm2}
\end{align}
The factor between braces is expressed as
$\lb\frac{\sqrt{l(m+1)}}{u+1}-1-\alpha\frac{1+C}{1+C\alpha}\frac{\sqrt{l(m+1)}}{u+1}\rb$.
The scaling factor $(\alpha D)^{m+l-1-2u}$ is maximum for $u=m$ where it
reaches $(\alpha D)^{p}$. We can consider up to bondary terms that
the $u$-sum goes only up to $m-1$. We can factorize $\alpha^{2}$ and prove that
the remaining terms are smaller than $1/\alpha$ on the ``mountain top'' or
$k$ outside this critical region.

For the $u$-sum, with $u$ running only to $m-1$ we have:
\begin{align}
  {\cal D} = i&\lbt\frac{\sqrt{1-\alpha}}{1+C\alpha}\rbt^{m+l-1}
  (1+C\alpha)^{-h-1}\sum_{u=0}^{m-1}(\alpha
  D)^{m+l-1-2u}{\cal A}(u;m,l-1,h+1)\nonumber\\
  &\times\lb\frac{\sqrt{l(m+1)}}{u+1}-1-\alpha\frac{1+C}{1+C\alpha}
  \frac{\sqrt{l(m+1)}}{u+1}\rb,\\
  {\cal D}\les&K\alpha^{2}\lbt\frac{\sqrt{1-\alpha}}{1+C\alpha}\rbt^{m+l-3}
  (1+C\alpha)^{-h-1}\nonumber\\
  &\times\sum_{u=0}^{m-1}(\alpha D)^{m+l-3-2u}{\cal A}(u;m-1,l-2,h+1)\nonumber\\
  &\times\frac{\sqrt{m(m+h+1)(l-1)(l+h)}}{(m-u)(l-1-u)}
  \lb\frac{\sqrt{l(m+1)}}{u+1}-1-c_{1}\alpha\frac{\sqrt{l(m+1)}}{u+1}\rb\nonumber\\
  = &K  \sum_{u=0}^{m-1} \Gamma (u,m-1,l-2, h+1)  E(u, m,l,h)
\end{align}   
where    $c_1 =\frac{1+C}{1+C\alpha} $, and
\begin{eqnarray}   
E (u, m,l,h) = \alpha^{2} \frac{\sqrt{m(m+h+1)(l-1)(l+h)}}{(m-u)(l-1-u)(u+1)(1+ C \alpha)}
\times\nonumber\\   
\lb(1- \alpha) \sqrt{l (m+1)} -(1+ C\alpha )  (u+1) \rb\nonumber .
\end{eqnarray}
Once more the subsequent procedure depends on the value of the indices
in the configuration space $\N^{3}$. If they stand on the mountain top, we pass to the variables
$x = v/k = C\alpha /(1+C\alpha )  + \delta x $, $  y = h+1 /k  = 1/(1+C)  + \delta y$, 
$z = p/k = \delta z$. We can assume $\alpha k\ges 1$ and we can evaluate 
$E$ as:
\begin{eqnarray}
E &\les &   O(1) [  (1- \alpha)  \sqrt{(1-y) (1-y-z)}  -   (1+ C \alpha) (1-x-y-z) ]    +   O(1/k) 
\nonumber\\ 
& \les & O( \vert \alpha \delta y \vert  + \vert \delta x \vert  + \vert \delta z \vert )  +   O(1/k) 
\label{numtermmountop}
\end{eqnarray}
The rest of the sum recontstructs the usual mountain bound of lemma \ref{goodbound1}.
Using the Hessian estimates $\delta y  \simeq 1/ \sqrt{\alpha k}$, $\delta x \simeq \delta z \simeq \sqrt {\alpha k} $, 
the $E$ correction simply provides an additional
factor $\sqrt {\alpha /k}$ to the estimate of Lemma \ref{goodbound1}. 
Adding the $\sqrt{m} \les \sqrt{k}$ factor we recover the desired $\sqrt{\alpha}$ factor.

If the indices are outside the critical region,
\begin{eqnarray}
{\cal D}& \les&  K\alpha^{2}\Gamma_{m-1,m+h;l+h-1,l-2} \max_{u\les
    m-1}\lbt\frac{\sqrt{m(m+h+1)(l-1)(l+h)}}{(m-u)(l-1-u)}\right.\\
&\times&\left.\lb\frac{l-1-u}{u+1}-c_{1}\alpha\frac{\sqrt{l(m+1)}}{u+1}\rb\rbt \les K\alpha^{2}\Gamma_{m-1,m+h;l+h-1,l-2}\,k\sqrt{1+\frac pm}\nonumber\\
 && \sqrt{m+1}\ {\cal D}\ \les \  K\sqrt\alpha\, e^{-c(p+\alpha k)}
\label{numtermouttop}
\end{eqnarray}
This is a better bound.\qed

\subsubsection{Boundary Terms}
\label{subsubboundary}
For purists we add the treatement of the boounary terms, and show that they do not perturb the previous computations. We start with the $u_{1}=0$ term in (\ref{boundaryterm1}). We have:
\begin{align}
  {\cal O}\defi&\lbt\frac{\sqrt{1-\alpha}}{1+C\alpha}\rbt^{m+1+l}(1+C\alpha)^{-h}(\alpha
  D)^{m+1+l}{\cal A}(0;m+1,l,h)\nonumber\\
  =&\lbt\frac{\sqrt{1-\alpha}}{1+C\alpha}\rbt^{m+1+l}(1+C\alpha)^{-h}(\alpha
  D)^{m+1+l}\sqrt{\binom{m+1+h}{m+1}\binom{l+h}{l}}\nonumber\\ 
  \les&(\alpha D)^{m+1+l}e^{-\alpha(C+1/2)(m+1+l)-\frac{C'}{2}
  \alpha h}\max_{h}e^{-\frac{C'}{4}\alpha
    h}\frac{(m+1+h)^{(m+1)/2}}{\sqrt{(m+1)!}}\nonumber\\
  &\hspace{1cm}\times\max_{h}e^{-\frac{C'}{4}\alpha h}\frac{(l+h)^{l/2}}{\sqrt{l!}}\\
  \les&\alpha^{(m+1+l)/2}\left(\frac{D\sqrt
      2}{\sqrt{C'}}\right)^{m+1+l}e^{-\alpha(C+1/2-\frac{C'}{4})(m+1+l)-\frac{C'}{2}\alpha
    h}\nonumber\\
  \les&(9\alpha)^{(m+1+l)/2}e^{-\frac 12\alpha(m+1+l)-\frac{C'}{2}
  \alpha h}\les(9\alpha)^{p/2}e^{-c\alpha k}
\end{align}
Then $\sqrt{m+1}{\cal O}\les K\sqrt\alpha\,e^{-c(\alpha k+p)}$ which is the
desired result.\\
We now study the $u=m$ term in (\ref{bondaryterm2})and prove that it obeys (\ref{eq:fullprop-bound}):
\begin{align}
  {\cal M}\defi&\lbt\frac{\sqrt{1-\alpha}}{1+C\alpha}\rbt^{m+l-1}(1+C\alpha)^{-h-1}(\alpha
  D)^{p}{\cal A}(m;m,l-1,h+1)\nonumber\\
  &\times\lb\lbt\frac{\sqrt{1-\alpha}}{1+C\alpha}\rbt^{2}(1+C\alpha)
  \frac{\sqrt{l(m+1)}}{m+1}-1\rb\nonumber\\
  \les&\,\Gamma_{m,m+h+1;l+h,l-1}\times\lb\frac{\sqrt{l(m+1)}}{m+1}-1-
  \alpha\frac{1+C}{1+C\alpha}\frac{\sqrt{l(m+1)}}{m+1}\rb\nonumber\\
  \les&\,\Gamma_{m,m+h+1;l+h,l-1}\times\lb\frac{l}{m+1}-1-
  \alpha\frac{1+C}{1+C\alpha}\sqrt{1+\frac{p}{m+1}}\rb\nonumber\\
  \les&\,K\Gamma_{m,m+h+1;l+h,l-1}\times\lb\frac{p}{m+1}+\alpha\rb.\label{eq:um3}
\end{align}
Remark that the $u=m$ term is not on the mountain top so that for the function
$\Gamma_{m,m+h+1;l+h,l-1}$ appearing in the equations (\ref{eq:um3}) we use
the bounds outside the mountain top region. Then we have to treat two
different cases according to the value of $\alpha k$. If $\alpha k\les 1$,
thanks to (\ref{eq:kpetit1}) 
\begin{align}
  {\cal M}\les& e^{-c\alpha k}\lbt\frac{\alpha
    \sqrt{(m+p)(m+1+p+h)}}{2}\rbt^{p}\times\lb\frac{p}{m+1}+\alpha\rb,\nonumber\\
  \sqrt{m+1}{\cal M}\les&\alpha p\sqrt{\frac{(m+p)(m+1+p+h)}{m+1}}\lbt\frac{\alpha
    k}{2}\rbt^{p-1}e^{-c\alpha k}\nonumber\\
  &+\alpha\sqrt{m+1}e^{-c\alpha k-c'p}\nonumber\\
  \les&\lbt\alpha p\sqrt{k}\sqrt{1+\frac{p}{m+1}}+\sqrt\alpha\rbt
  e^{-c(p+\alpha k)}\les K\sqrt\alpha e^{-c(p+\alpha k)}.
\end{align}
If $\alpha k\ges 1$,
\begin{align}
  \sqrt{m+1}{\cal M}\les&\lbt\frac{p}{\sqrt{m+1}}+\alpha\sqrt{m+1}\rbt
  e^{-c(p+\alpha k)}\les K\sqrt\alpha e^{-\frac c2(p+\alpha k)}.
\end{align}
\bigskip\\

\appendix

\section{The ordinary B=0 bound}
\setcounter{equation}{0}
\label{app1}

In this section we use the above analysis to revisit
the bounds on the $\phi^4_4$ propagator given in \cite{Rivasseau2005bh}.
By convention the indices of the $\phi^{4}$ propagator are two-dimensionnal indices.
Using obvious notations
\begin{align}
  G^{\phi^{4}}_{m,m+h;l+h,l}&=\int_{0}^{1}d\alpha\,
  \frac{(1-\alpha)^{\frac{\mu^{2}\theta}{8\Omega}}}{(1+C\alpha)^{2}}\,
  G^{\alpha,\phi^{4}}_{m,m+h;l+h,l},\label{eq:propphi4}\\
  G^{\alpha,\phi^{4}}_{m,m+h;l+h,l}&=\Gamma^{\alpha}_{m,m+h;l+h,l}\,
  (1-\alpha)^{h/2}\les\Gamma^{\alpha}_{m,m+h;l+h,l}\,e^{-\frac{\alpha
      h}{2}}.
\end{align}
We have then the following result (recalling $p=l-m$)
\begin{theorem}\label{thm:phi4}
The $\phi^4_4$ propagator in a slice obeys the bound
\begin{equation}\label{eq:boundphi4}  
    G^i_{m,n;k,l}\les K M^{-i}\min\lbt 1,(\alpha k)^{p}\rbt e^{-c(\alpha k+p)}
  \end{equation}
  for some (large) constant $K$ and (small) constant $c$ which depend only on $\Omega$.
\end{theorem}
{\proof} In this context the analysis of Lemmas \ref{lemma:boundakpetit} and
\ref{goodbound1} leads to the modified bound
\begin{align}
  G^\alpha_{m,n;k,l}&\les K e^{-\frac{\alpha h}{2}}\lbt\frac{\exp\{-\frac{c}{\alpha k}p^2  
    - \frac{c\alpha}{k} (h - \frac{k}{1+C})^2\}}{\sqrt{\alpha k}}  
  + \min\lbt 1,(\alpha k)^{p}\rbt e^{-c(\alpha k+p)}\rbt.\label{eq:phi4proof1}
\end{align}
The second term is already of the desired form in (\ref{eq:boundphi4}). Moreover:
\begin{align}
  \exp\lb-\frac{c\alpha}{k}\lbt h - \frac{k}{1+C}\rbt^2-\frac{\alpha
    h}{2}\rb&\les\exp\lb -\frac{c}{1+C}\alpha k-\alpha h\lbt \frac 12-\frac{2c}{1+C}\rbt\rb.
\end{align}
Then choosing $c$ small enough, the first term in (\ref{eq:phi4proof1}) is bounded by
\begin{align}
  K \exp\lb -c\alpha k-\frac{c}{\alpha k}p^{2}\rb&\les K\exp\lb -\frac
  c2\alpha k -c\sqrt2\,p\rb.
\end{align}
This completes the proof.\qed
\bigskip\\
Remark that in contrast with the Gross-Neveu case, the $\phi^{4}$ propagator has
no critical point in index space. The bound is the same for all $(m,l,h)\in\N^{3}$ and looks
like the bound (\ref{goodlargefieldbound}). Note also that the bound (\ref{eq:boundphi4}) allows to recover the
propositions 1, 2, 3 and 4 of \cite{Rivasseau2005bh} in a very direct manner.

\section{Proof of Lemma \ref{FermioXspace}}
\setcounter{equation}{0}
\label{app2}

 Let us compute first
\begin{eqnarray}
  Q^{-1}&=&\int_{0}^{\infty}dt\, e^{-tQ},\nonumber\\
  Q^{-1}(x,y)&=&\frac{\Omega}{\theta\pi}\int_{0}^{\infty}dt\frac{e^{-t\mu^{2}}}{\sinh(2\Ot t)}\,
  e^{-\frac{\Ot}{2}\coth(2\Ot t)(x-y)^{2}+\imath\Ot x\wedge y}e^{-2\imath\Ot
    t\gamma^{0}\gamma^{1}}\nonumber\\
  &\defi&\frac{\Omega}{\theta\pi}\int_{0}^{\infty}dt\,e^{-t\mu^{2}}e^{-2\imath\Ot
    t\gamma^{0}\gamma^{1}}\Gamma^t(x,y), \nonumber\\
\Gamma^t(x,y)  &=&
\frac{1}{\sinh(2\Ot t)}\,
e^{-\frac{\Ot}{2}\coth(2\Ot t)(x-y)^{2}+\imath\Ot x\wedge y}
\end{eqnarray}
with $\Ot=\frac{2\Omega}{\theta}$ and $x\wedge y=x^{0}y^{1}-x^{1}y^{0}$.\\

We only have to check that $e^{-tQ}$ is a solution of 
\begin{equation}
  \label{eq:expQ}
  \frac{dP}{dt}+QP=0.
\end{equation}
The constant is fixed by the requirement that in the limit $\Omega\to 0$, $Q^{-1}$
goes to the usual heat kernel.
\begin{eqnarray}
  \frac{d\Gamma^t}{dt}&=&\frac{e^{-\frac{\Ot}{2}\coth(2\Ot t)(x-y)^{2}
  +\imath\Omega x\wedge y}}{\sinh(2\Ot t)}\lbt-\frac{2\Ot\cosh(2\Ot t)}{\sinh(2\Ot
    t)}+\frac{\Ot^{2}}{\sinh^{2}(2\Ot t)}(x-y)^{2}\rbt,\nonumber\\
  \partial^{\nu}\Gamma^t&=&\frac{e^{-\frac{\Ot}{2}\coth(2\Ot
      t)(x-y)^{2}-\imath\Omega x\cdot\yt}}{\sinh(2\Ot t)}\lbt-\Ot\coth(2\Ot
  t)(x-y)^{\nu}-\imath\Omega\yt^{\nu}\rbt,\nonumber\\
  \Delta\Gamma^t&=&\frac{e^{-\frac{\Ot}{2}\coth(2\Ot t)(x-y)^{2}-\imath\Omega
      x\cdot\yt}}{\sinh(2\Ot t)}\times\nonumber\\
  &&\lbt -2\Ot\coth(2\Ot t)+\lbt -\Ot\coth(2\Ot
  t)(x-y)^{\mu}-\imath\Omega\yt^{\mu}\rbt^{2}\rbt\nonumber\\
  &=&\frac{e^{-\frac{\Ot}{2}\coth(2\Ot t)(x-y)^{2}-\imath\Omega
      x\cdot\yt}}{\sinh(2\Ot t)}\lbt -2\Ot\coth(2\Ot t)+\Ot^{2}\coth^{2}(2\Ot
  t)(x-y)^{2}\right.\nonumber\\
  &&\left.-\Omega^{2}\yt^{2}+ 2\imath\Omega\Ot\coth(2\Ot t)x\cdot\yt\rbt. 
\end{eqnarray}
The operator $L_{2}=x^{0}p_{1}-x^{1}p_{0}=-\imath x^{0}\partial_{1}+\imath
x^{1}\partial_{0}$ acting on $\Gamma^t$ gives:
\begin{eqnarray}
  L_{2}\Gamma^t&=&-\imath\frac{x^{0}}{\sinh(2\Ot t)}\lbt -\Ot\coth(2\Ot
  t)(x-y)^{1}-\imath\Omega\yt^{1}\rbt
  \nonumber\\
  &+&\imath\frac{x^{1}}{\sinh(2\Ot t)}\lbt -\Ot\coth(2\Ot
  t)(x-y)^{0}-\imath\Omega\yt^{0}\rbt
  e^{-\frac{\Ot}{2}\coth(2\Ot t)(x-y)^{2}-\imath\Omega
    x\cdot\yt}
  \nonumber\\
  &=&\frac{e^{-\frac{\Ot}{2}\coth(2\Ot t)(x-y)^{2}-\imath\Omega
    x\cdot\yt}}{\sinh(2\Ot t)}\lbt\imath\coth(2\Ot t)\Omega x\cdot\yt-\Ot x\cdot
  y\rbt. 
\end{eqnarray}
\begin{eqnarray}
  \lbt-\Delta+2\Ot L_{2}+\Ot^{2}x^{2}\rbt\Gamma^t&=&\frac{e^{-\frac{\Ot}{2}\coth(2\Ot t)(x-y)^{2}-\imath\Omega
    x\cdot\yt}}{\sinh(2\Ot t)}\lbt2\Ot\coth(2\Ot t)\right.\nonumber\\ 
&&\hspace{-2cm}-\Ot^{2}\coth^{2}(2\Ot  t)(x-y)^{2}+\Ot^{2}y^{2}
-2\imath\Omega\Ot\coth(2\Ot t)x\cdot\yt\nonumber\\
&&\hspace{-2cm}\left. +2\imath\Omega\Ot\coth(2\Ot t)x\cdot\yt-2\Ot^{2}x\cdot
  y+\Ot^{2}x^{2}\rbt\nonumber\\
  =\frac{e^{-\frac{\Ot}{2}\coth(2\Ot t)(x-y)^{2}
  -\imath\Omega  x\cdot\yt}}{\sinh(2\Ot t)}&\times&\big( 2\Ot\coth(2\Ot t)-\frac{\Ot^{2}}{\sinh^{2
    }(2\Ot t)}(x-y)^{2}\big)
\end{eqnarray}
With the $\mu^{2}$ and $\gamma^{0}\gamma^{1}$ terms, eq. (\ref{eq:expQ}) is satisfied.

For the full propagator it remains to compute
$e^{-2\imath\Ot t\gamma^{0}\gamma^{1}}$ and the action of
$\lbt-\ps+\mu-\Omega\xts\rbt$ on $e^{-tQ}$.
\begin{align}\label{expgamma01}
  e^{-2\imath\Ot t\gamma^{0}\gamma^{1}}&=\sum_{n\ges 0}\frac{(-2\imath\Ot
    t)^{2n}}{(2n)!}(-\mathds{1}_{2})^{n} + \sum_{n\ges 1}\frac{(-2\imath\Ot
    t)^{2n+1}}{(2n+1)!}(-\gamma^{0}\gamma^{1})^{n}\nonumber\\
  &=\cosh(2\Ot t)\mathds{1}_{2}-\imath\sinh(2\Ot t)\gamma^{0}\gamma^{1}.
\end{align}
With the convention $\slashed{p}=-\imath\ds$ we have
$-\slashed{p}=\imath\gamma^{\nu}\partial_{\nu}$ and
\begin{equation}
  -\slashed{p}\Gamma^t =\frac{\imath\gamma^{\nu}e^{-\frac{\Ot}{2}\coth(2\Ot t)(x-y)^{2}
  -\imath\Omega x\cdot\yt}}{\sinh(2\Ot t)}\big(-\Ot\coth(2\Ot
  -t)(x-y)_{\nu}-\imath\Omega\yt_{\nu}\big).
\end{equation}
which completes the proof of the Lemma.

\section{Proof of Lemma \ref{lemma:B0bis}}
\label{app3}
\setcounter{equation}{0}

We will restrict our attention to the case $h\ge 0$. We see that:
\begin{equation}
  \lim_{\alpha\to 0} {\cal E}=\delta_{mu}\delta_{lu}\rightarrow
  (1-\alpha)^{\frac{\theta}{8\Omega}H}|_{\alpha=O}=I
\end{equation}
so that we have the correct normalization.
One must now only check the differential equation:
\begin{equation}
  (1-\alpha)\frac{d}{d\alpha}(1-\alpha)^{\frac{\theta}{8\Omega}H}+
  \frac{\theta}{8\Omega}H(1-\alpha)^{\frac{\theta}{8\Omega}H}=0.
\end{equation}

A straightforward computation yields the result:
$$
-(1-\alpha)\frac{d}{d\alpha}[(1-\alpha)^{\frac{\theta}
  {8\Omega}H}]_{m,m+h;l+h,l}=\sum_{u=0}^{min(m,l)}{\cal A}(m,l,h,u)
\Big{(}C\frac{1+\Omega}{1-\Omega}\Big{)}^{m+l-2u}$$
$$\times\Big{[}
(2m+h+1)\Big{(}\frac{1}{2}+C\frac{1-\alpha}{1+C\alpha}\Big{)}-
(m-u)\Big{(}1+\frac{1-\alpha}{\alpha}+C\frac{1-\alpha}
{1+C\alpha}\Big{)}$$
\begin{equation}
  -(l-u)\Big{(}\frac{1-\alpha}{\alpha}-C\frac{1-\alpha}
  {1+C\alpha}\Big{)}
  \Big{]}{\cal E}(m,l,h,u).
\end{equation}
We will treat the last term in the above sum. Using the equality:
\begin{eqnarray}
  (l-u){\cal A}(m,l,h,u)&=&{\cal A}(m,l,h,u+1)\times \nonumber\\
  &&\hspace{-3cm}\Big{[}
  \frac{(m+1)(m+h+1)}{m-u}-(2m+h+1)+(m-u-1)
  \Big{]}
\end{eqnarray}
and changing the dummy variable from $u$ to $v=u+1$, the term 
rewrites as:
\begin{eqnarray}
  && \frac{\alpha C^2(\frac{1+\Omega}{1-\Omega})^2}{1+C\alpha}
  \sum_{v}
  \Big{[}
  \frac{(m+1)(m+h+1)}{m+1-v}-(2m+h+1)+(m-v)
  \Big{]}
  \nonumber\\   
  && \hspace{2cm} {\cal A}(m,l,h,v)
  \Big{(}C\frac{1+\Omega}{1-\Omega}\Big{)}^{m+l-2v}
  {\cal E}(m,l,h,v).
\end{eqnarray}
Coupling the identical terms in the two sums we get the coefficients 
of:
\begin{align} 
  2m+h+1&\rightarrow 
  \frac{1}{2}+C\frac{1-\alpha}{1+C\alpha}+
  \frac{\alpha C^2(\frac{1+\Omega}{1-\Omega})^2}{1+C\alpha}=
  C+\frac{1}{2}=\frac{1+\Omega^2}{4\Omega}\\
  m-u&\rightarrow (-1)
  \Big{[}\frac{1}{\alpha}+C\frac{1-\alpha}{1+C\alpha}+\frac{\alpha 
    C^2(\frac{1+\Omega}{1-\Omega})^2}{1+C\alpha}\Big{]}=
  -\frac{1+C\alpha}{\alpha}.
\end{align}
The complete sum is then:
\begin{eqnarray}
  &&\sum_{u}\Big{(}C\frac{1+\Omega}{1-\Omega}\Big{)}^{m+l-2u}
  {\cal A}(m,l,h,u){\cal E}(m,l,h,u)\times\\
  \nonumber
  &&\hspace{-.8cm} \Big{[}
  (2m+h+1)\frac{1+\Omega^2}{4\Omega}-(m-u)\frac{1+C\alpha}{\alpha}
  -\frac{(m+1)(m+h+1)}{m+1-u}\frac{\alpha 
    C^2(\frac{1+\Omega}{1-\Omega})^2}{1+C\alpha}\Big{]} .
\end{eqnarray}
Using 
\begin{eqnarray} (m-u){\cal A}(m,l,h,u)&=&\sqrt{m(m+h)}{\cal A}(m-1,l,h,u),\\
  \frac{(m+1)(m+h+1)}{(m+1-v)}{\cal A}(m,l,h,u)&=&
  \sqrt{(m+1)(m+h+1)}{\cal A}(m+1,l,h,u), \nonumber
\end{eqnarray}
one can cast the result into the form:
\begin{eqnarray}
  &&-(1-\alpha)\frac{d}{dt}[(1-\alpha)^{\frac{\theta}{8\Omega}H}
  ]_{m,m+h;l+h,l}=\\
  &&\frac{1+\Omega^2}{4\Omega}(2m+h+1)
  [(1-\alpha)^{\frac{\theta}{8\Omega}H}
  ]_{m,m+h;l+h,l}\nonumber\\
  &&-C\frac{1+\Omega}{1-\Omega}\sqrt{m(m+h)}
  [(1-\alpha)^{\frac{\theta}{8\Omega}H}]_{m-1,m-1+h;l+h,l}\nonumber\\
  &&-\sqrt{(m+1)(m+h+1)}C\frac{1-\Omega}{1+\Omega}
  [(1-\alpha)^{\frac{\theta}{8\Omega}H}]_{m+1,m+1+h;l+h,l} .
\end{eqnarray}
On the other hand:
\begin{eqnarray}
  &&\frac{\theta}{8\Omega}H_{m,m+h;p+h,p}=
  \frac{1+\Omega^2}{4\Omega}(2m+h+1)\delta_{mp}
  \nonumber\\
  &&\hspace{-.5cm} -\frac{1-\Omega^2}{4\Omega}
  [\sqrt{(m+h+1)(m+1)}~\delta_{m+1,p}+\sqrt{(m+h)m}~\delta_{m-1,p}]
\end{eqnarray}
and the differential equation is checked.

\paragraph{Acknowledgement}
We thank S.~Al~Jaber, V.~Gayral, J.~Magnen and R.~Wulkenhaar for discussions at
various stages of this work. We also thank our anonymous referee for
interesting comments which led to this improved version.

%\nocite{Seiberg:1999vs,MiRaSe,Chepelev:2000hm,Wilson:1973jj,Riv1,GrWu04-2,Langmann:2003cg,GrWu04-4,Rivasseau:2004az}
\bibliographystyle{utphys}
\bibliography{biblio-articles,biblio-books}

\end{document}